# Direct D-atom incorporation in radicals: An overlooked pathway for deuterium fractionation

Nureshan Dias[1], Ranil M. Gurusinghe[1], Bernadette M. Broderick[1], Tom J Millar[2,*], and Arthur G. Suits[1,*]

[1]Department of Chemistry, University of Missouri, Columbia MO 65211 USA

[2]School of Mathematics and Physics, Queen's University Belfast, University Road, Belfast BT7 1NN, UK

**Abstract**

Direct D-H exchange in radicals is investigated in a quasi-uniform flow employing chirped pulse mm-wave spectroscopy. Inspired by the H-atom catalyzed isomerization of $C_3H_2$ reported in our previous study, D atom reactions with the propargyl ($C_3H_3$) radical and its photoproducts were investigated. We observed very efficient D atom enrichment in the photoproducts through an analogous process of D addition/H elimination to $C_3H_2$ isomers occurring at 40K or below. Cyclic $C_3HD$ is the only deuterated isomer observed, consistent with the expected addition/elimination yielding the lowest energy product. The other expected addition/elimination product, deuterated propargyl, is not directly detected, although its presence is inferred by the observations in the latter part of the flow. There, in the high-density region of the flow, we observed both isotopomers of singly deuterated propyne attributed to stabilization of the H + $C_3H_2D$ or D + $C_3H_3$ adducts. The implications of these observations for the deuterium fractionation of hydrocarbon radicals in astrochemical environments is discussed with the support of a monodeuterated chemical kinetic model.

*Correspondence to: Tom.Millar@qub.ac.uk; suitsa@missouri.edu

**Introduction**

One of the fascinating observational results in astrochemistry has been the detection of deuterated molecules in high abundance. Many singly, doubly, and even a few triply deuterated molecules have been identified hitherto in various astronomical environments (Agúndez et al. 2021b; Lis et al. 2002; Parise et al. 2004; Parise et al. 2002; Roueff & Gerin 2003; Tiné et al. 2000; Turner 2001; Van der Tak et al. 2002; Vastel et al. 2003). Astonishingly, the observational evidence of the abundance ratios of these deuterated species relative to their fully hydrogenated forms are several orders of magnitude larger than expected based on the cosmic D/H ratio, $2 \times 10^{-5}$, set by Big Bang nucleosynthesis. For example, the abundance ratio of $ND_3/NH_3$ is around $10^{-3}$ - $10^{-4}$ in dark clouds, whereas the statistical value $(D/H)^3$ is roughly $3 \times 10^{-15}$, implying a $10^{11}$ - $10^{12}$ enhancement (Roueff et al. 2005). This is deuterium enrichment and occurs through a process called fractionation. The reactions that produce deuterium-enriched molecules in this process must selectively increase the deuterium content of product species. Deuterium forms slightly stronger bonds than hydrogen owing to small zero-point energy differences, and this can have a significant impact at low temperature. The effect of this on the reactive species in cold molecular clouds ensures that the deuterium is preferentially bonded into molecules.

Observation of deuterated molecules has also been used to probe various physical parameters such as temperature, ionization fraction, thermal history, and chemistry of astrophysical environments (Amano 2006; Brünken et al. 2014; Caselli 2002; Chen et al. 2010; Miettinen, Hennemann, & Linz 2011; Roberts & Millar 2000; Roberts & Millar 2007). However, the usefulness of deuterated molecules as a probe depends on our ability to understand the dynamics and kinetics of processes by which deuterium is incorporated into them. Considerable effort has thus been devoted to understanding deuterium fractionation in astronomical environments (Albertsson et al. 2013; Ceccarelli et al. 2014; Roberts, Herbst, & Millar 2002; Roberts & Millar 2000; Roueff & Gerin 2003; Willacy & Millar 1998). According to the current understanding, the formation of deuterated molecules in the gas phase is dominated by fast ion-neutral reactions (Dalgarno & Lepp 1984; Herbst 2003; Solomon & Woolf 1973; Watson 1977). Abundant molecular ions such as $H_3^+$, $CH_3^+$ and $C_2H_2^+$ undergo exchange reactions with HD, the reservoir for deuterium in dark clouds, to form deuterated ions $H_2D^+$, $CH_2D^+$ and $C_2HD^+$ (Roberts & Millar 2000). These exchange reactions are exothermic and favor the products as the D-bearing ions are more tightly bound due to their reduced zero-point energy. These ions can then transfer a deuteron to simple molecules such as $N_2$, CO and atomic oxygen to form simple deuterated molecular ions such as $N_2D^+$, $DCO^+$, $OD^+$ and initiate the processes leading to deuterium fractionation in many molecules in dark clouds.

The physical environment in cold dark clouds plays a major role in deuterium fractionation. In cold dense regions, where the central temperature of the cloud is 10 K or lower, abundant species which react with $H_2D^+$, such as CO, $N_2$ and O, are depleted by accretion on to the dust grains (Walmsley & Flower 2004), with large enhancements of deuterated species observed in sources with CO depletion (Bacmann et al. 2003; Punanova et al. 2016). Due to this depletion the enrichment of $H_2D^+$ is more pronounced (Roberts & Millar 2000) since under these conditions, $H_2D^+$ can further react with HD to form the other isotopologues of $H_3^+$ (Caselli, Sipilä, & Harju 2019; Roberts, Herbst, & Millar 2003). The detailed deuterium fractionation reactions from these isotopologues are described in studies by Roberts et al. and Flower et al. (Flower & Walmsley 2004; Roberts, et al. 2003; Walmsley & Flower 2004). In addition to gas-phase deuteration from $H_3^+$ isotopologues, grain surface chemistry is also important in deuterium fractionation (Caselli et al. 2002; Parise et al. 2006). Multiply deuterated species and some of the fully deuterated species are believed to form via surface reactions on dust grains (Brown & Millar 1989a). However, for some deuterated species the observations largely exceed the predictions and cannot be explained by these models.

In addition to these ion-molecule fractionation and surface reactions, there are very few neutral-atom reactions considered in the literature to form D-bearing molecules. One such significant pathway that involves neutral reactants is formation of OD in dark clouds. Croswell and Dalgarno were the first to predict the possibility of such reactions (Croswell & Dalgarno 1985). According to their predictions the exchange reaction, $D + OH \rightleftharpoons OD + H$, yields a large fractionation for OD in dark clouds. The kinetics of this exchange reaction was studied at room temperature and found to have a forward rate coefficient of $1.3 \times 10^{-10}$ cm$^3$ s$^{-1}$ (Margitan, Kaufman, & Anderson 1975). Croswell and Dalgarno adopted the same rate coefficient in their prediction and, given that the reaction is exothermic by 810 K, the reverse reaction is unimportant at low temperatures. They predicted OD/OH ratios to be around 0.01 or larger in dark clouds. Subsequently, Brown and Millar predicted that the OD/OH ratio can be large as unity in the clouds under freeze-out conditions (Brown & Millar 1989b). In these regions $D_3^+$ can form from through the successive reactions of $H_3^+$ with HD as discussed above. Under these extreme conditions $D_3^+$ can be more abundant than $H_3^+$ and the atomic D/H ratio can also be larger than unity in both gas phase and the grain surface resulting in enhanced OD fractionation. OD was first detected outside of the solar system in the cold low-mass protostar IRAS 16293-2422 using SOFIA, the Stratospheric Observatory for Infrared Astronomy. However, because OH was not observed with SOFIA at infrared wavelengths, an OD/OH ratio was not determined. Although observations at centimeter wavelengths can detect OH, they cannot be used to derive an accurate ratio since they sample an entirely distinct gas with a different density. Nevertheless, a very large fractionation for OD (OD/HDO ≈ 17-90) was inferred indirectly by Parise et al. (2012) comparing it with HDO in the same source.

Both c-$C_3H_2$ and $H_2CCC$ have been observed in dark cloud sources, including TMC-1, with a considerable abundance (Cox, Walmsley, & Güsten 1989; Kawaguchi et al. 1991). The third isomer, HCCCH, is expected to be found in dark clouds but has yet to be discovered due to its negligible dipole moment. The formation pathway for c-$C_3HD$ in dark clouds has been studied by several groups, and according to their work c-$C_3HD$ is primarily formed at low temperatures through the initiating reaction of c-$C_3H_2$ with $H_2D^+$ (Bell et al. 1988; Gerin et al. 1987; Spezzano et al. 2016). In extreme conditions, deuteration can also occur with $HD_2^+$ and $D_3^+$ forming the ionic intermediate $C_3H_2D^+$, followed by a dissociative recombination with electrons. In this paper, we demonstrate experimentally for the first time the formation of c-$C_3HD$ from direct D atom transfer reactions to the $C_3H_2$ radicals, and suggest this pathway may be of quite general application to the deuteration of hydrocarbon radicals and their subsequent reaction products in interstellar clouds.

In the present study, we apply broadband rotational spectroscopy in a low-temperature flow (Chirped-pulse/Uniform Flow, or CPUF) to show efficient D atom substitution for H atoms in $C_3H_2$ isomers and in propargyl radicals as well as H/D addition reactions to $C_3H_2D$ and $C_3H_3$. Although these latter addition reactions required collisional stabilization in our experiments to yield the detected products, we indicate them as radiative association reactions because it is the primary pathway for the formation of adducts in low density dark clouds. These reactions are summarized below:

D + C$_3$H$_2$ → c-C$_3$HD + H

D + C$_3$H$_3$ → CH$_2$CCD + H

D + C$_3$H$_3$ → CHDCCH + H

CH$_2$CCH + D → CH$_3$CCD + h𝜈

CH$_2$CCH + D → CH$_2$DCCH + h𝜈

CHDCCH + H → CH$_2$DCCH + h𝜈

CH$_2$CCD + H → CH$_3$CCD + h𝜈

We combine this with astrochemical modeling of various environments to explore the consequences for deuterium fractionation in these clouds. We argue that these pathways may be important in many other systems and contribute to deuterium fractionation under conditions where ionic pathways may be less important.

Previously we have studied the photodissociation of propargyl radical and quantified the product branching fractions for various C$_3$H$_2$ isomers at low temperature using CPUF in a quasi-uniform flow (Broderick et al. 2018). In that study we observed H atom catalyzed isomerization of C$_3$H$_2$ isomers to yield the more stable c-C$_3$H$_2$ isomer. c-C$_3$HD has been observed in many sources with fractionation ratios ranging from 0.08 to 0.23 (Chantzos et al. 2018), but current models underestimate its observed fractional abundance (Chantzos et al. 2018; Markwick, Charnley, & Millar 2001). This discrepancy, along with the results of our previous study, motivated us to incorporate D atoms to our experimental studies. Here, we report direct D atom incorporation to the C$_3$H$_2$ isomers in an analogous process. Following the recent discovery of propargyl radical in Taurus Molecular Cloud 1 (TMC-1), we suggest deuterated propargyl radicals will likely be abundant interstellar molecules (Agúndez et al. 2021a), although their detection is hampered by the small dipole moment of propargyl.

The deuterated isotopomer of propyne, $CH_2DCCH$, was first detected in TMC 1 by Gerin et al. (1992). About a decade later the other isotopomer, $CH_3CCD$, was detected in TMC 1 by Markwick et al. (2005). The main formation of these species involves ion-neutral reactions. In the present study we propose a possible pathway for the formation of deuterated propyne ($CH_2DCCH$ and $CH_3CCD$) in the interstellar medium by radiative association involving propargyl or its deuterated isotopologues. Based on the recent discovery of doubly deuterated propyne in L483 (Agúndez, et al. 2021b), we argue below that it can also be produced through radiative association reactions. Although experimentally we observe the products of these reactions, and the decay of some species, at this stage we cannot calibrate the densities as would be needed to measure the rates. Nevertheless, the experimental observations suggest the picture we outline, and the modeling we show provides confirmation of the potential impact. Concurrent with the experiments we performed astrochemical modelling using the UDfA Rate13 code and the results of the simulations are compared with observations from the literature.

**Experimental description**

The experiments are performed in our CPUF experimental setup described previously (Abeysekera et al. 2014; Oldham et al. 2014; Park & Field 2016). It comprises a pulsed Laval flow coupled to a chirped pulse mm-wave spectrometer. The CPUF experimental apparatus and the schematic of the experimental setup is depicted in Fig. 1(a) and (b). Here we use a "quasi-uniform" Laval flow in which the initial conditions are uniform and high density within the nozzle, which is followed by a second expansion to a low density and low temperature detection region which is optimized for the mm-wave signal (Broderick, et al. 2018; Dias et al. 2018). The number density of the uniform region is approximately few $10^{16}$ molecules/cm$^3$ and this persists for several centimeters. The density declines by several orders of magnitudes in the detection region preventing the collisional attenuation of the mm-wave signal. The characteristics of the quasi-uniform flow are described in depth in our earlier publication (Broderick, et al. 2018) and the details of current mm-wave circuitry are outlined in our previous publications (Abeysekera et al. 2015; Dias, Gurusinghe, & Suits 2022; Gurusinghe et al. 2021a).

In this experiment the flow consists of 1% of propargyl bromide (98% purity Sigma Aldrich; seeded in He) and 1% of ND$_3$ (99% purity Sigma Aldrich; seeded in He) for generation of radicals and D atoms. After the flow is established, 20 mJ of a loosely focused 193 nm laser radiation fires 10 μs before the first mm-wave pulse. Given the area of the laser beam this corresponds to a fluence of $6 \times 10^{16}$ photons cm$^{-2}$ per pulse. A single photon dissociates the propargyl bromide and ND$_3$ resulting mostly in propargyl (C$_3$H$_3$), ND$_2$, D atoms, and H atoms (Foley et al. 2019; Lee & Lin 1998; Nakajima et al. 1991). The C$_3$H$_3$ can react with ND$_2$ to give a range of products; these are the subject of a separate study reported elsewhere (Gurusinghe et al. 2021b). A second photon of the same wavelength that is absorbed within the 20 ns laser pulse may dissociate the propargyl radical giving mainly C$_3$H$_2$ isomers and H atoms. For each gas pulse, the rotational spectra are collected at 5 μs intervals up to 175 μs using the fast frame capabilities of the oscilloscope. This permits us to investigate the detailed dynamics of the reactions within the flow on this time scale.

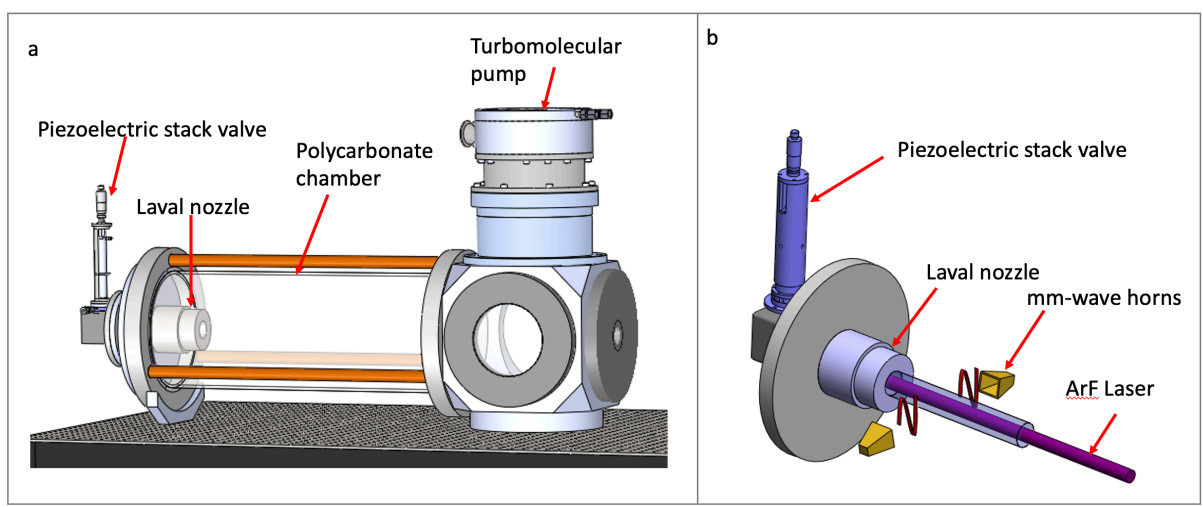

Figure 1. (a) Chirped pulse experimental apparatus. (b) Schematic of the experimental setup. To generate the He quasi-uniform flow, a piezoelectric stack valve generates a high throughput gas pulse, which is expanded through the Laval nozzle. The reaction is initiated by the ArF laser, which propagates counter to the flow. The products are detected by broadband rotational spectroscopy.

Broadband scans were used initially to identify the rotational lines corresponding to the products of photodissociation and reaction in the 70-92 GHz region. Upon identifying the lines, resonant frequency π/2 pulses of varying durations were used to excite the individual rotational lines observed in the initial broad chirps. Both up and down chirps were used to acquire and average the spectra to compensate for the dephasing effects. The rotational spectra corresponding to the photoproducts of propargyl radical dissociation and the products arising from the D atom reactions with radicals were obtained. The time evolution of the integrated intensities are then obtained from the corresponding rotational lines.

**Astrochemical modelling**

We have used the UDfA Rate13 codes (from now on Rate13 code, available at www.udfa.net ), with an updated, deuterated version of the UDfA Rate12 network to investigate the impact of inclusion of D-atom reactions on the abundance and fractionation of $c\text{-}C_3HD$ and other related species in our study (McElroy et al. 2013). The updated version of the reaction network now includes 846 species with 24952 reactions. New reactions involving propargyl, propyne and their deuterated analogues were added to this rate file and are given in the appendix with their collisional rate constants, and energy barriers. We used standard physical conditions for a cold, dark molecular cloud, similar to those often adopted for TMC-1, and for a warmer, denser cloud, similar to those in Orion Molecular Cloud (OMC) ridge. For TMC-1 we adopted a typical gas temperature of $T_g$ = 10 K and density of $4 \times 10^4$ cm$^{-3}$ (Cernicharo et al. 2021) and for OMC we use a gas temperature of 70 K and a slightly higher density, $1 \times 10^5$ cm$^{-3}$ (Sutton et al. 1995; Tercero et al. 2011) – although the latter gives 60K for $T_g$. We also used freeze on conditions to predict the fractional abundances in TMC-1 and chose the same density of $4 \times 10^4$ cm$^{-3}$ for these conditions. For both TMC-1 and OMC we used a standard cosmic ray ionization rate of 1.3 10$^{-17}$ s$^{-1}$. Relative to the total hydrogen nucleon density, we adopted initial fractional abundances of $1.5 \times 10^{-5}$ for HD which is similar to its cosmic abundance, and for atomic D we used $3 \times 10^{-10}$. We took initial elemental abundances from (McElroy, et al. 2013) but reduced their initial O abundance from $3.2 \times 10^{-4}$ to $1.3 \times 10^{-4}$ corresponding to a carbon-rich environment. Loomis et al. (2021) reported that a C/O ratio 1.1, similar to our value, gives the best fits for the observed species HC$_{11}$N in TMC-1. An initially low C/O ratio leads to a relatively low abundances of long carbo-chain and related species (Agúndez, et al. 2021a; Agúndez & Wakelam 2013).

We note that these simple one-point models neglect the intrinsic physical complexity of TMC-1 and the OMC region. For example, chemistry in a much more physically-detailed model of TMC-1 has been studied recently by Wakelam et al. (2021). Indeed, our models are also somewhat simplistic from a chemical modeling perspective since they also neglect the extensive spin-state chemistry involved in the deuteration of $H_2$ and its related ions, specifically $H_2D^+$, $HD_2^+$ and $D_3^+$, as well as the possibility that fractionation can also take place through grain surface reactions.

In particular, the efficacy of fractionation via these deuterated ions depends on the ortho-para ratio of $H_2$ since para-$H_2$ lies 170.5 K higher in energy and therefore possesses a much lower energy barrier in reactions that destroy its deuterated counterparts. Although radiative transitions between ortho and para forms are forbidden, proton exchange reactions with $H^+$ and $H_3^+$ are known to reduce the ortho-para ratio with time, from an initial (high-temperature) value of 3 to values around $10^{-3}$ at long times for a dark cloud similar to TMC-1, with detailed effects indicating a reduction in fractionation over models which neglect spin-state effects (Flower, Des Forêts, & Walmsley 2006; Majumdar et al. 2016). Nevertheless, we believe that there is some value in presenting these calculations to gain some estimate of the impact of such reactions on the fractionation of hydrocarbon molecules and to motivate future study.

**Experimental results**

Direct D atom exchange reactions were studied in a quasi-uniform Laval flow of helium in which precursor molecules were seeded then exposed to 193nm laser radiation as described above. In the past we have used the variation of the density in the quasi-uniform flow to characterize the density-dependent chemistry observed in the experiment, as regions of different conditions arrive at the detector at well-defined times (Broderick, et al. 2018; Gurusinghe, et al. 2021b). This approach is employed here as well. Rotational spectroscopy in the 70-92 GHz range is used to investigate the products formed as a function of delay between the laser initiation and the probe time (hence origin point in the flow). The results are given in Fig. 2 showing six spectral lines corresponding to rotational transitions of the six different products we observed. These correspond to l-$C_3H_2$, c-$C_3H_2$, c-$C_3HD$, $CH_3CCH$, $CH_2DCCH$ and $CH_3CCD$. The mm-wave transition frequencies and the line strength values are extracted from the Cologne Database for Molecular Spectroscopy (CDMS) and summarized in Table A1 in the appendix. The spectra are acquired in 5 μs intervals, and the laser fires between the first and second frame (5 μs after the first mm-wave pulse is fired). The lines that appear around 82.5 GHz in all frames are artifacts due to the mm-wave electronics and are independent of the valve and laser.

Figure (3.a) illustrates the time evolution plot of the total integrated intensities of all the products observed in this study normalized to the transition strength and the rotational partition function. Figure (3.b) depicts the products corresponding to the $C_3H_2$ isomers resulting from the direct photodissociation of the propargyl radical and the c-$C_3HD$ resulting from its reaction with D atoms. c-$C_3H_2$ appears 60 $\mu$s after the laser and persists until 120 $\mu$s. l-$C_3H_2$ appears after 30 $\mu$s and vanishes around 80 $\mu$s. The initial delay in the appearance of these two products is attributed to the distinct levels of rotational and vibrational cooling that the molecules undergo subsequent to their formation. Notable features in the time evolution plot are the disappearance of the l-$C_3H_2$, the appearance of c-$C_3HD$, the sharp rise in c-$C_3H_2$ at

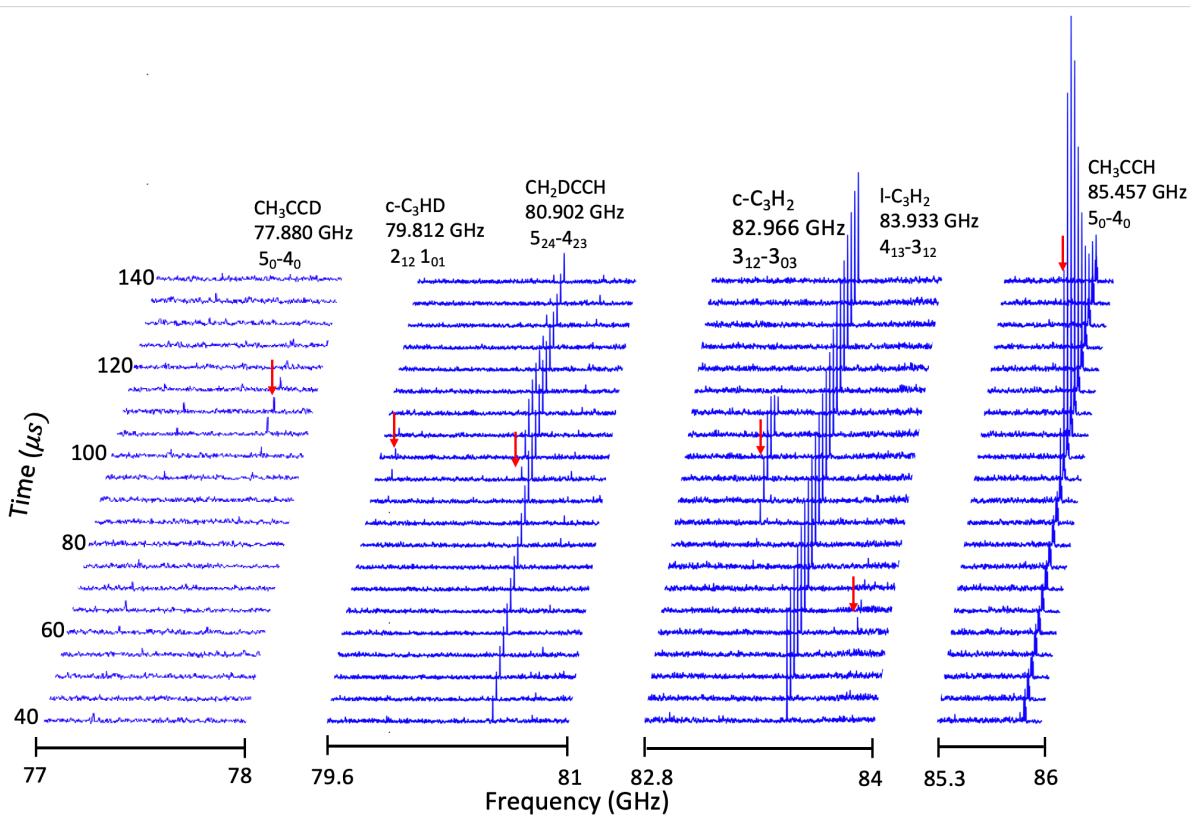

Figure 2. Rotational spectra of the products observed in this experiment. Probe delay after laser pulse is indicated on left. The red arrows indicate the positions of each rotational line observed, and the details of each line are indicated just above it.

around 75 $\mu$s and the rise in propyne and deuterated propyne signal at around 90 $\mu$s. Figure (3.c) shows the time evolution plot of propyne and its deuterated isotopologues resulting from the reaction of deuterated propargyl with H atoms or normal propargyl with D atoms. These products appear later in the spectra corresponding to the high-density region near the nozzle throat. The propyne products appear 90 $\mu$s after the laser and peak at 135 $\mu$s. In these spectra the total integrated intensity for $CH_2DCCH$ is roughly twice as high as that for $CH_3CCD$, and both species exhibit similar time evolution.

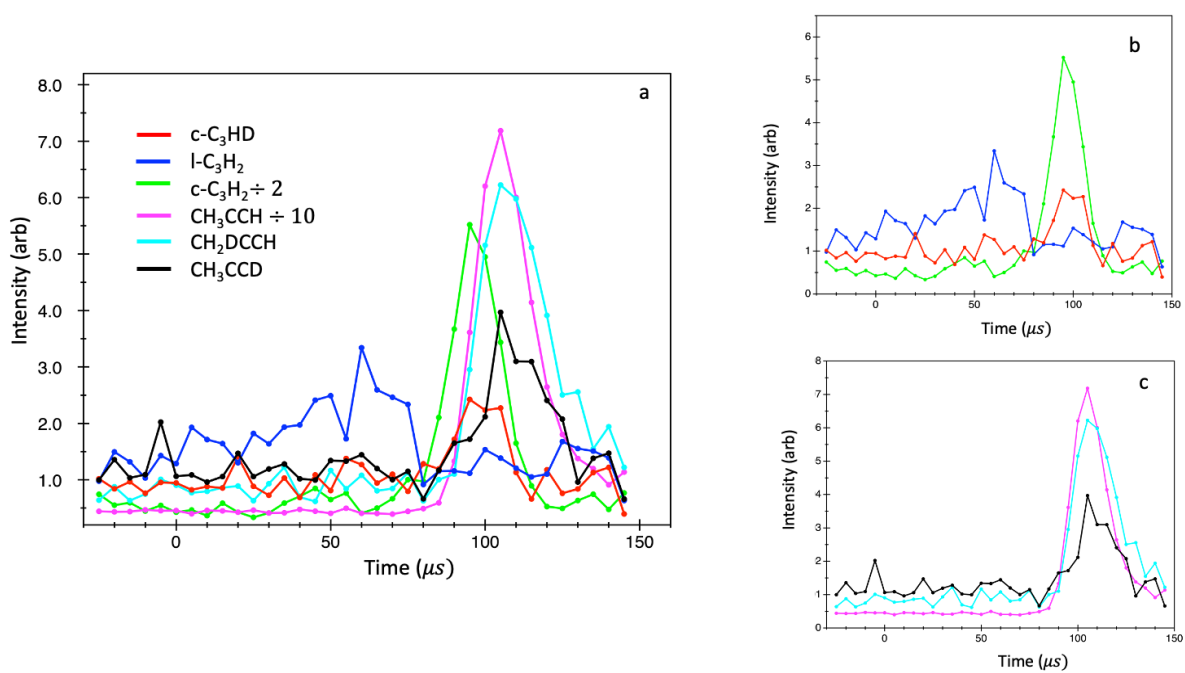

Figure 3(a) The time evolution integrated intensities of products observed in the experiment (b) time evolution integrated intensity plot of $C_3H_2$ species observed and (c) the time evolution integrated intensity plot of propyne species observed in this study.

**Experiments: Discussion**

We first discuss the D-H exchange results seen in the experiment. Although to our knowledge D-H exchange reactions between $C_3H_n$ (n=2,3) isomers have never been explored before, the addition/elimination of H atoms to these species has been investigated using both experimental and theoretical methods. Klippenstein et al. (2015) have employed high-level electronic structure methods with classical trajectory calculations to determine the microcanonical rate constants for the dissociation of propargyl radical and the reverse, recombination of $C_3H_2$ isomers with H atoms under a variety of conditions (Klippenstein, Miller, & Jasper 2015). The results of the electronic structure methods, as well as high-pressure rate constants for H atom addition reactions with $C_3H_2$ isomers at temperatures ranging from 500 to 2500 K, demonstrate that this recombination is fast and proceeds without a barrier. Our previous experimental study shows that these reactions are also fast at low temperatures (Broderick, et al. 2018). In the flow, the propargyl produced by this recombination is much colder than the initial propargyl formed via photodissociation. However, the propargyl formed through this recombination still possesses some energy and requires a third body to stabilize it through collision or it will redissociate. Such stabilization is not possible in our medium density region. Therefore, it dissociates back to $C_3H_2$ but now exclusively to c-$C_3H_2$, the most stable isomer, in a process of H-catalyzed isomerization:

H + $C_3H_2$ → $C_3H_3$* → c-$C_3H_2$ + H

The importance of these H atom catalyzed isomerization reactions in Titan's atmosphere is also discussed by Hébrard et al. (2013).

We note that other species are present in the flow as discussed in our previous work (Gurusinghe 2021b, Broderick 2018) and mentioned above. Aside from deuterium atoms, the primary deuterated species will be $ND_2$. Reaction of $ND_2$ with $C_3H_2$ or $C_3H_3$ to give any of the detected species is endothermic and not likely to contribute to the measured signals under our experimental conditions.

The present study confirms the efficient incorporation of D atoms into $C_3H_2$ isomers resulting from the reaction of propargyl bromide photoproducts with D atoms, and it thus reveals a new pathway to deuterium fractionation. In a manner analogous to H + $C_3H_2$ recombination, D + $C_3H_2$ recombination is fast and barrierless and mimics the H atom catalyzed isomerization. The combination of $C_3H_2$ with D atoms forms deuterated propargyl which will dissociate exclusively to c-$C_3HD$ in our lower or medium density region at low temperature. We examined for l-$C_3HD$ in and found no lines that corresponded to it in our spectra, confirming that the dissociation was exclusive to c-$C_3HD$. We observe the appearance of c-$C_3HD$ around 110 $\mu$s which corresponds to the medium density region in the flow. In this region, termolecular collisions are negligible, and the products formed in this region cannot be stabilized by collisions.

We also observed propyne and its deuterated isotopologues appearing around 100 $\mu$s, which corresponds to the higher density region where products can be stabilized through third-body collisions. This must result either from H addition to deuterated propargyl radicals, or D addition to normal propargyl radicals, or both. Deuterated propargyl may be formed in the through D-H exchange just as we describe for c-$C_3$HD formation. However, its small dipole moment precludes monitoring propargyl directly in this study. Harding et al. (2007) have shown H recombination with propargyl is also fast and barrierless. Different reactivities are predicted for the two resonance structures of the propargyl radical (i.e., when the attack occurs on the methylenic or "head" side or the acetylenic "tail" side), but they appear to approach similar rates at low temperature. In any case, scrambling of the H/D location is possible, in particular via cyclic intermediates c-$C_3H_2$D or c-$C_3H_3$D, for both pathways that may lead to deuterated propyne.

In Fig. 3 we see that the intensity for the deuterated propyne isotopomer, $CH_2$DCCH, is roughly twice that for $CH_3$CCD supporting the importance of scrambling en route to formation of deuterated propyne. However, we observed different abundance ratios for $CH_2$DCCH to $CH_3$CCD depending on experimental conditions and propargyl radical precursor. In some cases, the intensities of both deuterated propyne isotopomers were nearly identical, whereas in others, $CH_3$CCD was not detected. The variation in relative yield we observe suggests competition between quenching of the hot propyne or allene adducts vs. isomerization and scrambling, with final branching depending strongly on precise pressure and H and D concentrations. This will be the subject of a future study.

Both reaction pathways identified here are strongly exoergic for D-H exchange:

D + $C_3H_2$ → c-$C_3HD$ + H + 943 K

D + $C_3H_3$ → $CH_2CCD$ + H + 948 K

D + $C_3H_3$ → CHDCCH + H + 900 K.

This contrasts with the ion-neutral reaction that initiates most of the deuterium fractionation as previously understood:

$H_3^+$ + HD → $H_2D^+$ + $H_2$ + 230 K

The consequence of this is that deuterium fractionation via the D-H exchange mechanism can contribute at significantly higher temperatures than $H_2D^+$ as discussed in the modeling below.

**Astrochemical modeling results**

*TMC-1*

The time evolution of the fractionation, $X_{(D)}/X_{(H)}$ of c-$C_3HD$, CHDCCH, $CH_2CCD$, $CH_2DCCH$, and $CH_3CCD$ and was modeled using the Rate13 code and their abundances are compared for cases with and without the D-H atom exchange reactions. These results, in terms of the time evolution of their molecular fractionation ratios, are summarized for TMC-1 in Fig 4(a) with freezeout conditions off and in 4b with freezeout conditions on. Initial abundances and the conditions used in the modeling are described in the astrochemical modelling section. Fig. 4(a) shows that the inclusion of D exchange reactions to the $C_3H_2$ isomers significantly increases the fractionation of c-$C_3HD$, to a level slightly higher than its observed value in TMC-1 (see Table 1 for a comparison of observed and predicted fractional abundances and fractionation ratios at a time of $1.5 \times 10^5$ years.) Moreover, the exchange reaction of D atoms with the propargyl radical

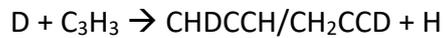
D + C$_3$H$_3$ → CHDCCH/CH$_2$CCD + H

also increases the fractionation of deuterated propargyl radicals. The fractionation for deuterated propyne, which forms primarily from the radiative associations of H atoms with CHDCCH and CH2CCD, increases slightly as the consequence of the increased deuterated propargyl fractionation. Table 2 summarizes the reactions that contribute more than 10% to the formation and destruction of species mentioned above for the freeze on and freeze off models. Our analysis shows that the main reactions that lead to the formation of c-C$_3$HD, and CHDCCH/CH$_2$CCD are the D atom transfer reactions to the C$_3$H$_2$ and CH$_2$CCH radicals, respectively. Fig. 4(b) shows the time evolution of the fractionation of related species in TMC-1 when we include the freeze-out conditions. For this, we chose $1.5 \times 10^5$ years as the time when our model most closely reproduces the fractionation results. This condition further increased the fractionation of above-mentioned species. Under these conditions, the fractionation of c-C$_3$HD becomes slightly larger than its observed and projected values in the absence of freezeout conditions. Note that although the fractionation ratios are large at long times, ~ 10$^6$ years, the underlying molecular abundances are very small due to freeze-out.

Table 1 Observed and predicted fractional abundances and fractionation ratios in TMC1 with freeze out off and freeze out on conditions

| | | | c-$C_3$HD | CHDCCH | $CH_2$CCD | $CH_2$DCCH | $CH_3$CCD |
|---|---|---|---|---|---|---|---|
| Without freeze out conditions $1.5 \times 10^5$ year and number density of 50000 $cm^{-3}$ | Abundance | Observed | 1.20E-09[a] | - | - | 9.20E-10[b] | 2.00E-10[c] |
| | | Without D reactions | 2.59E-09 | 3.56E-12 | 1.78E-12 | 3.76E-10 | 1.44E-10 |
| | | With D atom reactions | 7.40E-08 | 6.25E-12 | 3.12E-12 | 5.05E-10 | 2.07E-10 |
| | Fractionation | Observed | 0.08-0.16[a] | - | - | 0.106[b] | 0.04-0.11[c] |
| | | without D reactions | 0.007 | 0.007 | 0.003 | 0.005 | 0.002 |
| | | With D atom reactions | 0.271 | 0.014 | 0.007 | 0.007 | 0.003 |
| With freeze out conditions $1.5 \times 10^5$ year and number density of 50000 $cm^{-3}$ | Abundance | Without D reactions | 2.35E-09 | 1.35E-12 | 6.74E-12 | 1.96E-10 | 7.15E-11 |
| | | With D atom reactions | 1.24E-08 | 2.55E-12 | 1.27E-12 | 2.74E-10 | 1.04E-10 |
| | Fractionation | Without D reactions | 0.049 | 0.018 | 0.009 | 0.020 | 0.007 |
| | | With D atom reactions | 0.334 | 0.035 | 0.017 | 0.028 | 0.011 |

Notes [a] Chantoz et al. (2018), [b] (Cabezas et al. 2021a), [c] Markwick et al. (2005).

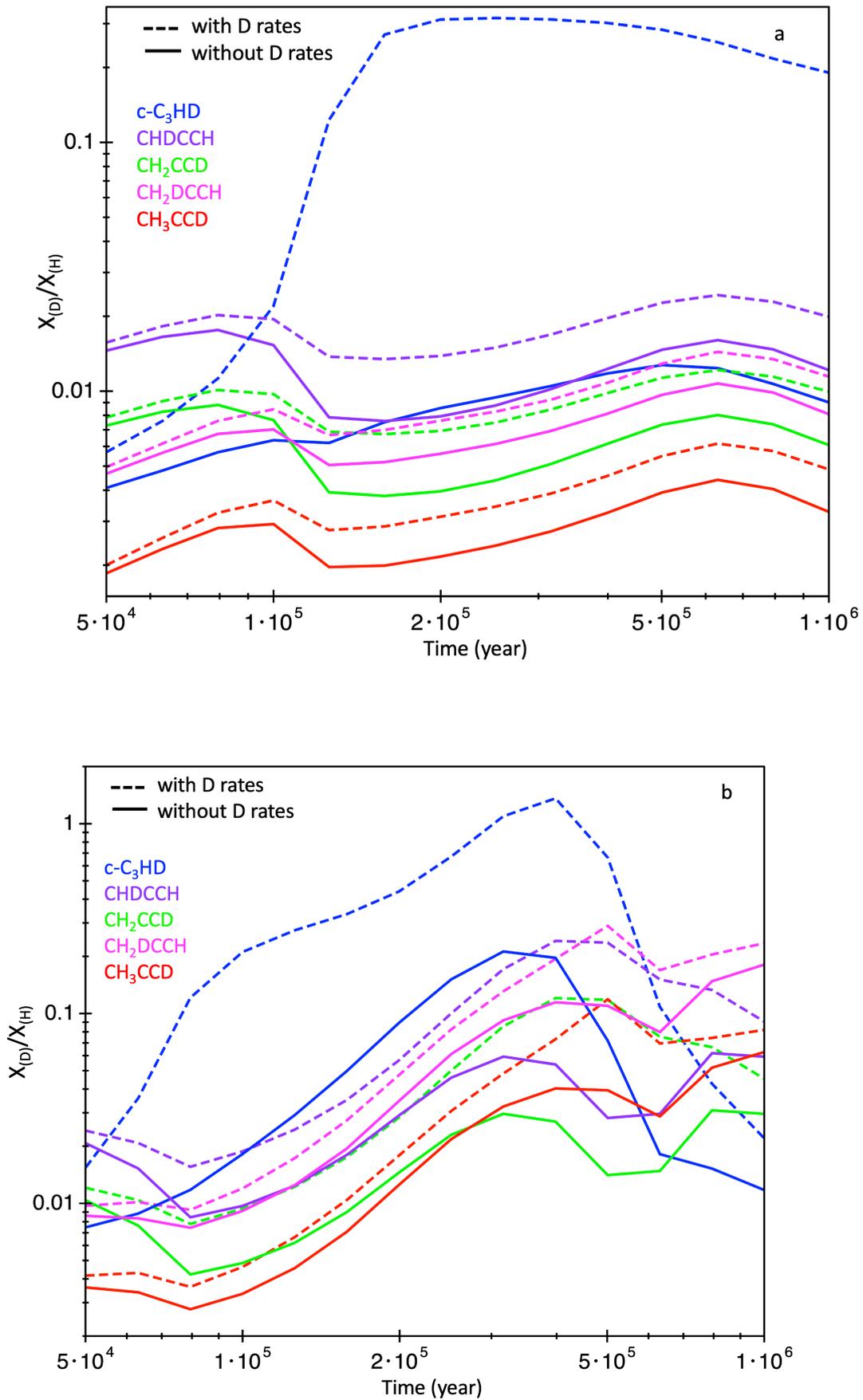

Figure 4. Modeled fractionation of species of interest in TMC-1 (a) without freezeout and (b) with freezeout conditions.

Table 2 Main formation and destruction reactions of species of interest in TMC-1 at a model time of $1.5 \times 10^5$ yr.

| Species | Main Formation reaction | % | Main destruction reaction | % |
|---|---|---|---|---|
| c-$C_3$HD | $D + C_3H_2 \rightarrow C_3HD + H$ | 78 | $HCO^+ + C_3HD \rightarrow C_3H_2D^+ + CO$ | 33 |
|  | $e^- + C_3H_2D^+ \rightarrow C_3HD + H$ | 21 |  |  |
| CHDCCH | $D + CH_2CCH \rightarrow CHDCCH + H$ | 38 | $CHDCCH + H \rightarrow CH_2DCCH + PHTN$ | 38 |
|  | $C + C_2H_3D \rightarrow CHDCCH + H$ | 35 | $CHDCCH + H \rightarrow CH_2CCHD + PHTN$ | 38 |
| $CH_2CCD$ | $D + CH_2CCH \rightarrow CH_2CCD + H$ | 38 | $CH_2CCD + H \rightarrow CH_3CCD + PHTN$ | 38 |
|  | $C + C_2H_3D \rightarrow CH_2CCD + H$ | 35 | $CH_2CCD + H \rightarrow CH_2CCHD + PHTN$ | 38 |
| $CH_2DCCH$ | $CH_2CCH + D \rightarrow CH_2DCCH + PHTN$ | 11 | $CH_2DCCH + CN \rightarrow CH_2DC_3N + H$ | 22 |
|  | $e^- + C_3H_6D^+ \rightarrow CH_2DCCH + H_2 + H$ | 33 | $CH_2DCCH + CN \rightarrow CH_3C_3N + D$ | 16 |
|  | $CHDCCH + H \rightarrow CH_2DCCH + PHTN$ | 51 | $CH_2DCCH + C \rightarrow C_4H_2D + H$ | 27 |
| $CH_3CCD$ | $CH_2CCD + H \rightarrow CH_3CCD + PHTN$ | 47 | $CH_3CCD + CN \rightarrow CH_3C_3N + D$ | 10 |
|  | $e^- + C_3H_6D^+ \rightarrow CH_3CCD + H_2 + H$ | 27 | $CH_3CCD + CN \rightarrow CH_2DC_3N + H$ | 27 |
|  | $CH_2CCH + D \rightarrow CH_3CCD + PHTN$ | 10 | $CH_3CCD + C \rightarrow C_4H_2D + H$ | 29 |

*Orion Ridge Cloud*

Figure 5 shows a comparison of the predicted fractionation of species mentioned above in the warmer and denser Orion ridge cloud. At higher temperatures, above 20-25 K, the back reaction of $H_2$ with $H_2D^+$ becomes efficient and deuteration via this ion becomes unimportant. However, at such temperatures the D atom incorporation to the radicals is still fast and barrierless and the reverse reaction prohibited. As in the case for TMC-1, when we include the D atom incorporation to the radicals, the fractionation of all the species mentioned above increased. However, the comparison between predicted and observed fractionations of c-$C_3$HD and deuterated propargyl is not possible in OMC due to the lack of observational evidence for the presence of both deuterated species. Table 3 compares the observed and predicted fractional abundances and fractionation ratios at a time of $2 \times 10^5$ years. According to our findings, the D atom transfer reactions to the $C_3H_2$ and $CH_2CCH$ radicals are the primary pathway for the production of both c-$C_3$HD and deuterated propargyl (see Table 4).

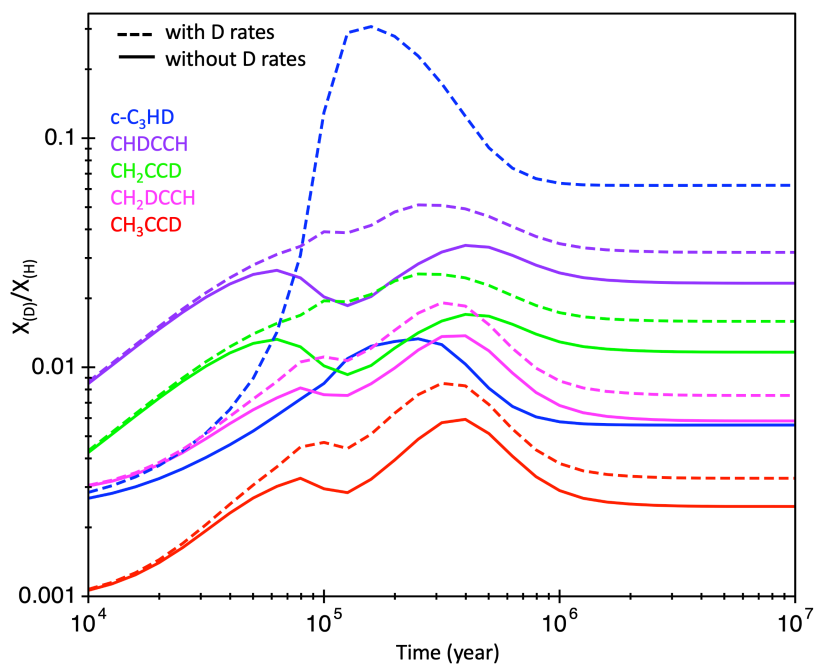

Figure 5. Modeled fractionation of species of interest in Orion Molecular Cloud.

Table 3 Observed and predicted fractional abundances and fractionation ratios in OMC at a model time of $2 \times 10^5$ yr.

|  |  | c-$C_3$HD | CHDCCH | $CH_2$CCD | $CH_2$DCCH | $CH_3$CCD |
|---|---|---|---|---|---|---|
| Abundance | Without D reactions | 1.62E-09 | 1.22E-11 | 6.11E-12 | 1.38E-10 | 5.48E-11 |
|  | With D atom reactions | 2.74E-08 | 2.30E-11 | 1.15E-11 | 2.00E-10 | 8.69E-11 |
| Fractionation | Observed | - | - | - | <0.05[a] | - |
|  | without D reactions | 0.012 | 0.024 | 0.012 | 0.010 | 0.004 |
|  | With D atom reactions | 0.278 | 0.047 | 0.024 | 0.015 | 0.006 |

Notes [a] (Millar 1997).

*Table 4. Main formation and destruction reactions of species of interest in OMC at a model time of 2 x 10$^5$ yr.*

| Species | Main Formation reaction | % | Main destruction reaction | % |
|---|---|---|---|---|
| c-C$_3$HD | D + C$_3$H$_2$ → C$_3$HD + H | 69 | HCO$^+$ + C$_3$HD → C$_3$H$_2$D$^+$ + CO | 48 |
| | e- + C$_3$H$_2$D$^+$ → C$_3$HD + H | 29 | | |
| CHDCCH | D + CH$_2$CCH → CHDCCH + H | 46 | CHDCCH + H → CH$_2$DCCH + PHTN | 35 |
| | C + C$_2$H$_3$D → CHDCCH + H | 38 | CHDCCH + H → CH$_2$CCHD + PHTN | 35 |
| CH$_2$CCD | D + CH$_2$CCH → CH$_2$CCD + H | 46 | CH$_2$CCD + H → CH$_3$CCD + PHTN | 35 |
| | C + C$_2$H$_3$D → CH$_2$CCD + H | 38 | CH$_2$CCD + H → CH$_2$CCHD + PHTN | 35 |
| CH$_2$DCCH | CH$_2$CCH + D → CH$_2$DCCH + PHTN | 11 | CH$_2$DCCH + CN → CH$_2$DC$_3$N + H | 49 |
| | e- + C$_3$H$_6$D$^+$ → CH$_2$DCCH + H$_2$ + H | 24 | CH$_2$DCCH + CN → CH$_3$C$_3$N + D | 16 |
| | CHDCCH + H → CH$_2$DCCH + PHTN | 51 | | |
| CH$_3$CCD | CH$_2$CCD + H → CH$_3$CCD + PHTN | 59 | CH$_3$CCD + CN → CH$_3$C$_3$N + D | 16 |
| | e- + C$_3$H$_6$D$^+$ → CH$_3$CCD + H$_2$ + H | 18 | CH$_3$CCD + CN → CH$_2$DC$_3$N + H | 49 |
| | CH$_2$CCH + D → CH$_3$CCD + PHTN | 13 | | |

**Astrochemical modelling discussion**

The principal pathways to deuterium fractionation are well established and believed to occur chiefly through ion-molecule reactions and grain surface chemistry. However, for some species the observations vastly exceed what simulations predict. This inconsistency has been observed for many molecules including NH$_2$D, DCN, N$_2$D$^+$, c-C$_3$HD and DC$_3$N. For some of these species the modeled fractionation becomes somewhat consistent with the observed value at the steady state. However, many of these dark cloud sources have not yet reached the steady state.

For c-C$_3$HD, our basic model, without D-atom exchange reactions and freeze out, gives $X_D/X_H$ = 0.007 at 1.5 x 10$^5$ years and 0.015 at steady-state, much lower than the range observed in TMC-1, 0.08-0.16. Some of these observational values can be uncertain due to large optical depths or self-absorption in the non-deuterated species. However, c-$C_3$HD has been observed abundantly in many dark cloud sources, indicating that there are few uncertainties associated with its measurements (Chantzos, et al. 2018). Interestingly, in some sources the fractionation for c-$C_3$HD becomes comparable or higher than its ostensible deuterated source, $H_2D^+$ (0.1) or $CH_2D^+$ (0.02). This is an indication that the deuteration of c-$C_3$HD may not occur exclusively from the commonly expected reactions. Howe and Millar have tested possible alternative routes to the deuterium fractionation (Howe & Millar 1993). Direct deuteration of ionic species by reaction with D atoms or HD was discussed. Fractionation for c-$C_3$HD improved with the inclusion of the reaction of $C_3H_3^+$ with HD followed by the dissociative recombination. Millar et al. (1989) also noted this discrepancy and suggested that the enhanced ratio may be due to preferential ejection of H atoms from the ionic intermediate during the dissociative recombination (Millar, Bennett, & Herbst 1989).

Our modelling indicates that more than 78% of c-$C_3$HD is formed through the D atom exchange involving the $C_3H_2$ radicals in cold conditions, and in warm conditions such as in the Orion cloud, 69% of c-$C_3$HD is formed through the same reaction (see Table 4). The fractionation ratio increased to 0.27 under TMC-1-like conditions by including D at om transfer reactions to the $C_3H_2$ isomers in the model, comparable to the observed range of 0.08-0.16. The calculated value increases to 0.33 when freeze-out is applied. In warm OMC-like conditions, the fractionation has increased from 0.012 to 0.28 when the D-atom exchange reaction is included. However, since c-$C_3$HD has not yet been detected in this source, a comparison between theory and observation is not possible at present.

As shown in figures 4 and 5, our chemical modeling predicts increased abundances for deuterated propargyl when we include D atom exchange reactions, increasing from 0.007 to 0.014 for CHDCCH and from 0.003 to 0.007 for $CH_2CCD$ in cold TMC-1-like conditions. In warm conditions the fractionation increased from 0.024 to 0.047 for CHDCCH and from 0.012 to 0.024 for $CH_2CCD$. Both our experiments and modeling indicate that deuterated propargyl should be present at an observable level in both TMC-1 and the Orion molecular cloud. Agúndez et al. (2021a) has recently discovered the propargyl radical in TMC-1 using the Yebes 40m telescope with a fractional abundance relative to $H_2$ of 8.7 x $10^{-9}$ suggesting that the propargyl radical is one of the most abundant radicals ever found in TMC-1. (Agúndez, et al. 2021a) This further implies deuterated isotopologues of propargyl should also present in these regions. Searches for these species in TMC-1 are thus urgently required to test the deuteration mechanism discussed herein.

The radiative association reaction: $\mathrm{H + CH_2CCH \rightarrow CH_3CCH + h\nu}$, is included in the KIDA database (Kinetics Database for Astrochemistry) (Wakelam et al. 2012). We have added the corresponding D atom reactions in our chemical modeling (See Table 1 in appendix). According to our model, these radiative association reactions between $\mathrm{D + CH_2CCH}$/ CHDCCH are primarily responsible for the generation of deuterated propyne. In cold TMC-1 like conditions, the fractionation increased from 0.005 to 0.007 for $CH_2DCCH$ and from 0.002 to 0.003 for $CH_3CCD$. However, the fractionation predicted by the model is still significantly lower than the observed fractionation (0.106 for $CH_2DCCH$ and 0.04-0.11 for $CH_3CCD$). With the introduction of freeze out conditions, the prediction further increases toward the observed values. Under these conditions the fractionation of $CH_2DCCH$ and $CH_3CCD$ is 0.028 and 0.011, respectively, but much higher at later times ($\approx 5 \times 10^5$ years). At these times, the fractional abundances are still significantly higher, but lower than the observed values. The plot of fractional abundances of all the species observed in TMC-1 was calculated by incorporating the D atom reactions and freeze out conditions, is shown in Fig.1 in the appendix. In warm (OMC) conditions the fractionation improved from 0.010 to 0.015 for $CH_2DCCH$ and from 0.004 to 0.006 for $CH_3CCD$. Here the predicted value for $CH_2DCCH$ is around 4 times less than the observed value (0.026), showing the importance of D atom transfer reactions in warm temperatures. Because $CH_3CCD$ has yet to be observed in OMC, no comparison is made here. Deuterated allene should also form in our flow, in the same manner as deuterated propyne, however due to its vanishing dipole moment, it is not possible to observe it in our instrument.

Doubly deuterated cyclopropenylidene; c-$C_3D_2$ is observed in nearby dark clouds and more recently doubly deuterated propyne has been discovered in the dense core L483 (Agúndez, et al. 2021b; Spezzano et al. 2013). These species have a significant fractionation in these cold sources, about a few percent that of the monodeuterated species. Our results indicate that the reactions, $D + C_3HD \rightarrow C_3D_2 + H$ and $D + CH_2CCD\ (CHDCCH) \rightarrow CH_2DCCD(CHD_2CCH) + h\nu$ can also occur in dark clouds. We looked carefully for these species in our flow but, we did not see any evidence for them.

Cabezas et al. (2021b) have recently reported detection of $CH_2DC_4H$ (but not $CH_3C_4D$) in TMC-1, and considered two possible routes to its formation (Cabezas et al. 2021b). One involved $C_2D$ reaction with propyne, the other $C_2H$ reaction with $CH_2DCCH$. When these reactions were assumed to proceed directly rather than with full H(D) scrambling, little deuterium enrichment in $CH_2DC_4H$ was found. Interestingly, their alternative model invoked a D-H exchange reaction involving propyne, D + CH$_3$CCH -> CH$_2$DCCH + H as part of the mechanism to account for the formation of deuterated propyne as a precursor to the observed species. Reasonable agreement with the observed fractionation for CH$_2$DC$_4$H was achieved. However, the authors assumed that the reaction was fast and barrierless, but theory and experiment both support a barrier of ~11 kJ/mol making the reaction negligible at low temperature (Wang, Hou, & Gu 2000). Barriers such as this or larger are typical for H atom reaction with closed shell hydrocarbons, but not with radicals. Cabezas et al. (2021b) also assumed that reaction at the acetylenic site was 1/3 the rate at the allenic site for statistical reasons to help account for the absence of the CH$_3$C$_4$D species. We suggest analogous D/H exchange involving radicals should be considered. As mentioned above, with freeze-out conditions, the radiative association involving D + propargyl brings the model closer to the observed abundances of deuterated propynes that serve as a precursor to CH$_2$DC$_4$H. Moreover, although our current ratios are still too low, a larger rate coefficient for the D + propargyl reaction could reconcile model and observation. We have taken equal rates for all RA reactions involving D and H atoms, but this is not necessarily the case.

**Astrochemical implications**

It is apparent from figure 4 that the chemical model underestimates the deuterium fractionation of species of interest when we disregard the D atom reactions. However, when we include the D/H exchange with radicals it improves the fractionation for c-C$_3$HD and deuterated propynes and increases that for deuterated propargyl. In cold conditions such as TMC-1 the modeled fractionation of c-C$_3$HD becomes 0.270, slightly higher than the observed value. For the deuterated propargyl the fractionation increased by a factor of 1.5 and becomes close to one percent that of the fully hydrogenated species. The fractionation of c-$C_3$HD increases to 0.334 in cold conditions with the inclusion of D atom reactions and a freeze out condition. In comparison to the no-D, no-freeze out conditions, this increases fractionation of c-$C_3$HD by a factor of $\approx$ 40. However, with this condition the fractionation exceeds the observed value. Also, with the freezeout on and the inclusion of D atom reactions, the fractionation for deuterated propargyl increases by a factor of 3.5. The CHDCCH and $CH_2$CCD are now 3 and 2 % of their H parent ($CH_2$CCH), respectively. Given the large abundance of propargyl detected in TMC-1, deuterated propargyl may be detectable in this object. According to our modeling, a significant amount of propargyl is destroyed by RA reactions, lowering their abundances. Our model predicts an increase in fractionation for deuterated propyne when we include the radiative association reactions, but the observed abundances in TMC-1 still greatly exceed the predictions. When freeze-out is included, our calculated values increase but still fall factors of 3-4 below those observed. However, the analysis of the formation and destruction reactions at $1.5 \times 10^5$ year shows that the radiative association of deuterated propargyl with H is the main formation pathway for deuterated propyne in cold conditions and we encourage further studies on this system to more accurately determine its rate coefficient. According to our calculations, the fractionation of $CH_2$DCCH is a factor of 2-3 higher than that of $CH_3$CCD. When the RA reactions are excluded, the fractionation ratio of $CH_2$DCCH/$CH_3$CCD does not change significantly, implying that the higher fractionation for $CH_2$DCCH is primarily due to the greater abundance of CHDCCH than $CH_2$CCD. In warm clouds, where the fractionation by $H_2D^+$ species is less significant, the fractionation of hydrocarbon species also increases indicating the importance of D + radical reactions. We find that our calculated value for the fractionation of $CH_2$DCCH is within a factor of two of that observed when D atom reactions are included in our model. However, to our knowledge, the majority of the deuterated species we have focused on here do not yet have abundances reported for warm clouds such as OMC.

**Conclusions**

In this paper, chirped pulse mmW spectroscopy in a quasi-uniform flow was used to investigate the direct D atom incorporation to $C_3H_2$ and propargyl radicals at low temperature. The detection of c-$C_3HD$ shows efficient D atom transfer to $C_3H_2$ radicals through D addition/H elimination. The deuterated propargyl radical, the other anticipated product, is not directly observed, although its presence is inferred from the deuterated propyne observed in the high-density region of the flow. The formation of deuterated propyne in the flow also suggests that it can be formed in dark clouds via radiative association reactions with H atoms. Our models also indicate that such radiative association reactions of H and D atoms with hydrocarbon radicals can contribute significantly to deuterium fractionation. Laboratory and theoretical studies are thus needed to determine the associated rate coefficients. Concurrent to experiments, the dark cloud modeling demonstrated an increased fractionation for all deuterated species in this study, particularly when freeze-out was included. This preliminary study demonstrates that D-H exchange reactions are an important and overlooked pathway for deuterium fractionation in astrochemical environments. We encourage modelers to incorporate such reactions in models of deuterium chemistry that are more detailed in terms of both physics and chemistry than those discussed here.

This work was supported by the NSF under award number CHE-1955239. TJM thanks the STFC for support through grants ST/P000312/1 and ST/T000198/1 and gratefully acknowledges the receipt of a Leverhulme Emeritus Fellowship.

**Appendix**

*mm-wave Data relevant to the species observed in this study*

Table A1. Rotational transition frequency, lower state energy, transition quantum numbers and line strength value of each species observed in this study.

| Species | Transition Frequency (GHz) | Lower state energy (cm$^{-1}$) | $J_{up}$-$J_{low}$ | Line strength value (S$\mu^2$) |
|---|---|---|---|---|
| CH$_3$CCD | 77.880 | 5.19 | $5_0$-$4_0$ | 5.89 |
| c-C$_3$HD | 79.812 | 1.40 | $2_{12}$-$1_{01}$ | 41.58 |
| CH$_2$DCCH | 80.902 | 5.39 | $5_{24}$-$4_{23}$ | 3.07 |
| c-C$_3$H$_2$ | 82.966 | 8.38 | $3_{12}$-$3_{03}$ | 31.33 |
| l-C$_3$H$_2$ | 83.933 | 13.48 | $4_{13}$-$3_{12}$ | 189.13 |
| CH$_3$CCH | 84.457 | 6.14 | $5_0$-$4_0$ | 6.14 |

**Further Astrochemical Modeling Details**

Table A2 Summarizes the new reactions added to the reaction network. The references for the rates are included in the table. The (*) indicates the values obtained from the corresponding non-deuterated reaction. When no experimental or theoretical data is available, the rate coefficients are estimated from corresponding high pressure H atom recombination reactions using reasonable assumptions based on statistical ratios (e.g. D + CH$_2$CCH, (Harding, Klippenstein, & Georgievskii 2007). For these back reactions, the barrier,

$\gamma$, is the difference in zero-point energy of the D and H species. However, since we calculate all models at low temperatures, once the barrier is greater than about **$6T_g$,** then the back reaction, i.e. the H atom reaction, becomes unimportant and other reactions such as with ions dominate loss of the deuterated species.

Table A2. The new rates added to the rate file. The rate coefficients in this table are expressed in the standard form: $k(T) = \alpha(T/300)^\beta \exp(-\frac{\gamma}{T})$ cm³ s⁻¹. All the barriers adopted ensure that the back reactions with H atoms are unimportant in setting deuterium fraction for the species considered here.

| Reaction type | Reaction | α | β | γ | Reference |
|---|---|---|---|---|---|
| NN | D + H$_2$CCC → C$_3$HD + H | 2.00E-10 | 0 | 0 | KIDA* |
| NN | H + C$_3$HD → H$_2$CCC + D | 2.00E-10 | 0 | 575 | KIDA* |
| NN | D + CH$_2$CCH → CHDCCH + H | 6.67E-11 | 0 | 0 | |
| NN | D + CH$_2$CCH → CH$_2$CCD + H | 3.33E-11 | 0 | 0 | |
| NN | H + CHDCCH → CH$_2$CCH + D | 1.00E-10 | 0 | 892 | |
| NN | H + CH$_2$CCD → CH$_2$CCH + D | 1.00E-10 | 0 | 865 | |
| NN | D + C$_3$H$_2$ → C$_3$HD + H | 4.00E-11 | 0 | 0 | |
| NN | H + C$_3$HD → C$_3$H$_2$ + D | 4.00E-11 | 0 | 936 | |
| NN | H + H$_2$CCC → C$_3$H$_2$ + H | 2.00E-10 | 0 | 0 | KIDA |
| NN | D + C$_2$H → C$_2$D + H | 5.00E-11 | 0 | 0 | |
| NN | D + HCO → DCO + H | 5.00E-11 | 0 | 0 | |
| NN | H + C$_2$D → C$_2$H + D | 5.00E-11 | 0 | 654 | |
| NN | H + DCO → HCO + D | 5.00E-11 | 0 | 630 | |
| NN | D + HCCCH → C$_3$HD + H | 1.00E-10 | 0 | 0 | KIDA* |
| NN | H + C$_3$HD → HCCCH + D | 1.00E-10 | 0 | 1300 | KIDA* |
| NN | H + HCCCD → C$_3$HD + H | 1.00E-10 | 0 | 0 | KIDA* |
| NN | H + C$_3$HD → HCCCD + H | 1.00E-10 | 0 | 2000 | KIDA* |
| NN | H + HCCCH → C$_3$H$_2$ + H | 1.00E-10 | 0 | 0 | KIDA |
| NN | H + C$_3$H$_2$ → HCCCH + H | 1.00E-10 | 0 | 2200 | KIDA |
| NN | H + HCCCD → C$_3$H$_2$ + D | 1.00E-10 | 0 | 0 | KIDA |
| NN | D + C$_3$H$_2$ → HCCCD + H | 1.00E-10 | 0 | 2900 | KIDA* |
| RA | H + CH$_2$CCH → CH$_3$CCH + PHOTON | 1.00E-13 | -1.5 | 0 | KIDA |
| RA | H + CH2CCH → CH$_2$CCH$_2$ + PHOTON | 1.00E-13 | -1.5 | 0 | KIDA |
| RA | H + CHDCCH → CH$_2$DCCH + PHOTON | 1.00E-13 | -1.5 | 0 | KIDA* |
| RA | H + CH$_2$CCD → CH$_3$CCD + PHOTON | 1.00E-13 | -1.5 | 0 | KIDA* |
| RA | D + CH$_2$CCH → CH$_2$CCHD + PHOTON | 1.00E-13 | -1.5 | 0 | KIDA* |
| RA | D + CH$_2$CCH → CH$_3$CCD + PHOTON | 3.33E-14 | -1.5 | 0 | KIDA* |
| RA | D + CH$_2$CCH → CH$_2$DCCH + PHOTON | 6.66E-14 | -1.5 | 0 | KIDA* |
| RA | H + C$_3$H$_2$$^+$ → C$_3$H$_3$$^+$ + PHOTON | 2.00e-15 | 0 | 0 | Ref a |
| RA | D + C$_3$H$_2$$^+$ → C$_3$H$_2$D$^+$ + PHOTON | 2.00e-15 | 0 | 0 | Ref a* |
| RA | D + C$_3$H$_2$$^+$ → CH$_2$CCD$^+$ + PHOTON | 6.66e-16 | 0 | 0 | Ref a* |
| RA | D + C$_3$H$_2$$^+$ → CHDCCH + PHOTON | 1.33e-15 | 0 | 0 | Ref a* |
| RA | H + CH$_2$CCD → CH$_2$CCHD + PHOTON | 1.00E-13 | -1.5 | 0 | KIDA* |
| RA | H + CHDCCH → CH$_2$CCHD + PHOTON | 1.00E-13 | -1.5 | 0 | KIDA* |
| DR | $C_3H_3^+ + e^- \rightarrow H_2CCC + H$ | 2.00e-08 | -0.50 | 0 | KIDA |
| DR | $C_3H_2D^+ + e^- \rightarrow HDCCC + H$ | 1.33e-08 | -0.5 | 0 | KIDA* |
| DR | $C_3H_2D^+ + e^- \rightarrow H_2CCC + D$ | 6.66e-09 | -0.5 | 0 | KIDA* |
| DR | $C_3H_3^+ + e^- \rightarrow HCCCH + H$ | 5.00e-08 | -0.5 | 0 | KIDA |
| DR | $C_3H_2D^+ e^- \rightarrow HCCCD + H$ | 3.33e-08 | -0.5 | 0 | KIDA* |
| DR | $C_3H_2D^+ + e^- \rightarrow HCCCH + D$ | 1.67e-08 | -0.5 | 0 | KIDA* |
| IN | $H^+ + CH_3CCH \rightarrow C_3H_3^+ + H_2$ | 7.50e-10 | -0.5 | 0 | KIDA |
| IN | $H^+ + CH_3CCH \rightarrow CH_2CCH^+ + H_2$ | 7.50e-10 | -0.5 | 0 | KIDA |
| IN | $H^+ + CH_2CCH_2 \rightarrow C_3H_3^+ + H_2$ | 7.50e-10 | 0 | 0 | KIDA |
| IN | $H^+ + CH_2CCH_2 \rightarrow CH_2CCH^+ + H_2$ | 7.50e-10 | 0 | 0 | KIDA |
| CE | $H^+ + CH_2CCH_2 \rightarrow C_3H_4^+ + H$ | 1.50e-9 | 0 | 0 | KIDA |
| IN | $He^+ + CH_3CCH \rightarrow C_3H_3^+ + He + H$ | 2.00e-10 | -0.5 | 0 | KIDA |
| IN | $He^+ + CH_3CCH \rightarrow CH_2CCH^+ + He + H$ | 2.00e-10 | -0.5 | 0 | KIDA |
| IN | $He^+ + CH_2CCH_2 \rightarrow C_3H_3^+ + He + H$ | 2.00e-10 | 0 | 0 | KIDA |
| IN | $He^+ + CH_2CCH_2 \rightarrow CH_2CCH^+ + He + H$ | 2.00e-10 | 0 | 0 | KIDA |

Notes [(a)] (Loison et al. 2017)

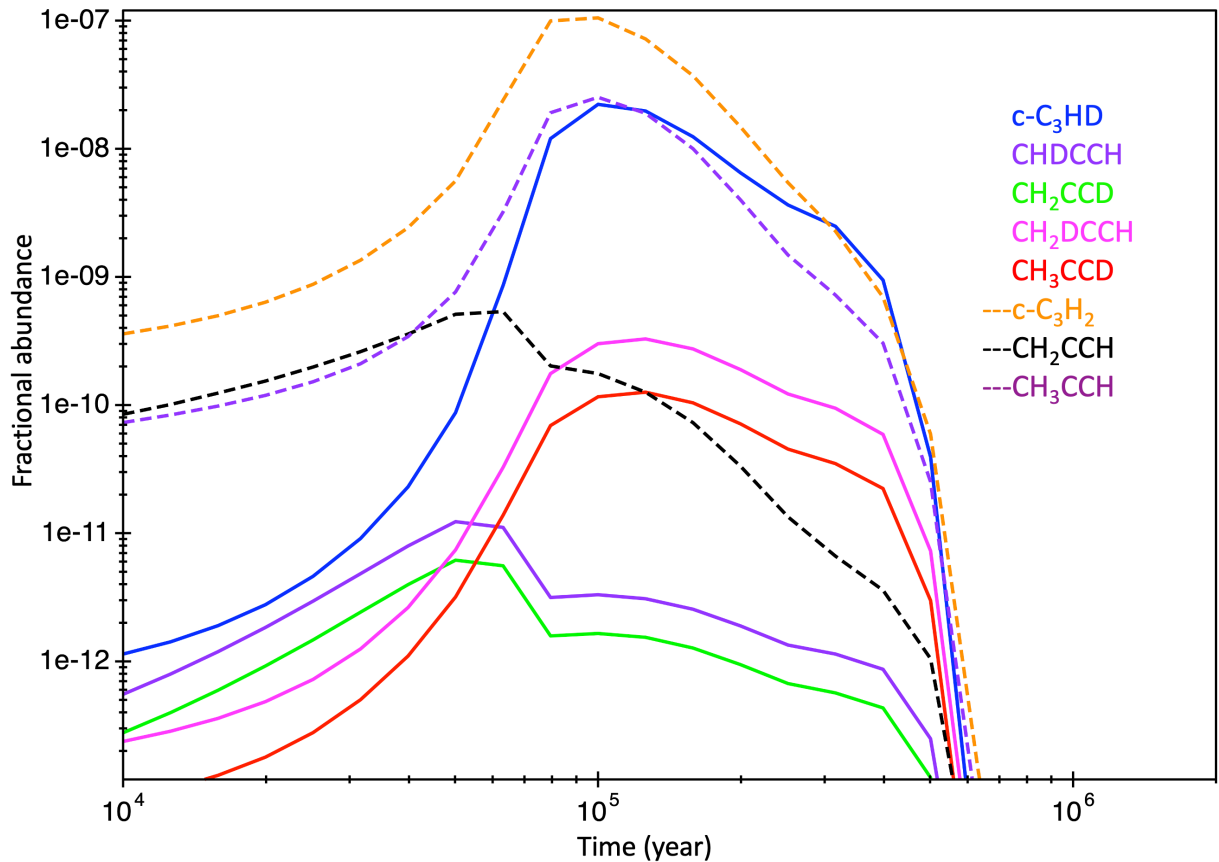

Figure A1. The fractional abundance of each species observed in this study was determined using UDfA Rate 13 codes and an updated, deuterated version of the UDfA Rate 12 network. The fractional abundance of each species is calculated using new D atom reactions and freeze out conditions. See main text for more details.